\documentclass[]{interact}

\usepackage{epstopdf}% To incorporate .eps illustrations using PDFLaTeX, etc.
\usepackage[caption=false]{subfig}% Support for small, `sub' figures and tables

\usepackage[numbers,sort&compress]{natbib}% Citation support using natbib.sty
\bibpunct[, ]{[}{]}{,}{n}{,}{,}% Citation support using natbib.sty
\makeatletter% @ becomes a letter
\def\NAT@def@citea{\def\@citea{\NAT@separator}}% Suppress spaces between citations using natbib.sty
\makeatother% @ becomes a symbol again

\theoremstyle{plain}% Theorem-like structures provided by amsthm.sty

\theoremstyle{definition}

\theoremstyle{remark}

\usepackage{hyperref}
\usepackage{listings}

\begin{document}

\articletype{ARTICLE TEMPLATE}% Specify the article type or omit as appropriate

\title{Modelling turbulence via numerical functional integration using Burgers' equation}

\author{
\name{I. Honkonen\textsuperscript{a}\thanks{CONTACT I. Honkonen. Email: ilja.honkonen@fmi.fi. ORCID: 0000-0002-9542-5866} and J. Honkonen\textsuperscript{b}}
\affil{\textsuperscript{a}Finnish Meteorological Institute, Helsinki, Finland; \textsuperscript{b}Finnish National Defence University, Helsinki, Finland}
}

\maketitle

\begin{abstract}
We investigate the feasibility of modelling turbulence via numeric functional integration.
By transforming the Burgers' equation into a functional integral we are able to calculate equal-time spatial correlation of system variables using standard methods of multidimensional integration.
In contrast to direct numerical simulation, our method allows for simple parallelization of the problem as the value of the integral within any region can be calculated separately from others.
Thus the calculations required for obtaining one correlation data set can be distributed to several supercomputers and/or the cloud simultaneously.

We present the mathematical background of our method and its numerical implementation.
We are interested in a steady state system with isotropic and homogeneous turbulence, for which we use a lattice version of the functional integral used in the perturbative analysis of stochastic transport equations.
The numeric implementation is composed of a fast serial program for evaluating the integral over a given volume and a parallel Python wrapper that divides the problem into subvolumes and distributes the work among available processes.
The code is available at \url{https://github.com/iljah/hdintegrator} for anyone to download, use, study, modify and redistribute.

We present velocity cross correlation for a 10x2 lattice in space and time respectively, and analyse the computational resources required for the integration.
We also discuss potential improvements to the presented method.
\end{abstract}


\section{Introduction}
\label{intro}

Understanding turbulence is likely relevant for phenomena of any scale, from particle collisions in an accelerator \cite{blaizot13} and human blood circulation \cite{sabbah76} to atmospheric and oceanic circulation, solar wind \cite{goldstein15} and even galaxy clusters \cite{zhuravleva14}.
Most commonly used methods for studying turbulence involve solving Navier-Stokes (NS) equations in various forms.
Implicit large eddy simulation (LES) methods do not include a term for viscosity in NS equation, but numerical errors due to finite accuracy of floating point numbers act implicitly as an artificial viscosity.
Explicit, or plain, LES methods include turbulent effects using averaged NS equations or with empirical small scale models.
Direct numerical simulation (DNS) methods add a viscosity term explicitly into NS equation which allows, in principle, to fully describe turbulent flow.
In practice DNS methods are compulationally very expensive if one is to describe both large and small spatial scales and their interaction, and require a powerful supercomputer.
The problem is exacerbated by the fact that, for example, doubling the Reynolds number increases the amount of memory required by at least an order of magnitude \cite{McComb}, making it currently impossible to model many systems with realistic Reynolds numbers.

We present a new approach for modelling turbulence via numerical functional integration.
By transforming the Burgers' equation into a functional integral, we are able to calculate equal-time spatial correlation of system variables using standard methods of multidimensional integration.
In contrast to direct numerical simulation, our method allows for simple parallelization of the problem as the value of the integral within any region can be calculated separately from others.
Thus the calculations required for obtaining one correlation data set can be distributed to several supercomputers and/or the cloud simultaneously.

\section{Mathematical background}
\label{math}

We use a lattice version of the functional integral used in the perturbative analysis of stochastic transport equations (see, e.g., \cite{turbo,Vasiliev}). We are considering the case of isotropic and homogeneous turbulence, therefore we use periodic boundary conditions in space. Since we are interested in the steady state of the system, we have imposed periodic condition with respect to time. Due to these choices the normalization constant for calculation of moments of the velocity field has to be calculated numerically.

The eventual goal is the random force uncorrelated in time but concentrated in a narrow slice of (small) wave numbers in the Fourier representation. In order to assess the feasibility of this approach, 
we are considering here the simplest case of random force uncorrelated both in space and time.

It is customary to analyze the statistical properties of the turbulent system in terms of single-time correlation functions.
Under these conditions the actual multi-dimensional integral to represent the functional integral for the two-point correlation function is of the form 
\begin{multline}
\label{Gen2}
\langle v(t_i,x_k)v(t_i,x_n)\rangle=C\int \prod_{l=1}^L \prod_{m=1}^M dv(t_l,x_m)\, v(t_i,x_k)v(t_i,x_n)\\
\times\exp\Biggl[-{1\over 2D\tau}\sum_{l=1}^L\sum_{m=1}^M\biggl(v(t_{l+1},x_m)-v(t_l,x_m)+\Biggl\{v(t_l,x_m)\left[{v(t_l,x_{m+1})-v(t_l,x_{m-1})\over 2a}\right]\\-
\nu\left[{v(t_l,x_{m+1})-2v(t_l,x_{m})+v(t_l,x_{m-1})\over a^2}\right]\Biggr\}\tau\biggr)^2 \Biggr]\,,
\end{multline}
where $C$ is the normalization constant. In (\ref{Gen2}) the lattice constant is 
$a=x_n-x_{n-1}=1$, the time step $\tau=t_{l}-t_{l-1}=1$, the (kinematic) viscosity is $\nu=1$ and the variance of the random force is $D=1$.
For economy of notation we have not written explicitly in (\ref{Gen2}) the periodic conditions imposed on the velocity field in time and space.

\section{Numerical implementation}
\label{numeric}

We use the HDIntegrator program \cite{honkonen17} for evaluating the functional integrals in parallel on the Finnish Meteorological Institute's Cray XC 40 supercomputer.
The integration volume is subdivided into smaller and smaller subvolumes until one or more of user-defined criteria for convergence is reached.
Convergence of the solution for each subregion is checked by evaluating the integral twice, where the second evaluation uses some factor of more samples decided by the user.
Currently the solution is defined as converged when one more of the following criteria are satisfied:
1) the absolute relative difference between results within a subvolume is smaller than some factor, or
2) the absolute difference between results is smaller than some value, or
3) the maximum absolute value of results is less than some value.
Listing \ref{nsphere} shows an example invocation and output of hdintegrator for calculating (half of) the volume of a unit sphere in 4 dimensions with a 3-dimensional integrand.

\lstset{language = bash, caption = {Example invocation and output of the parallel Python wrapper for calculating (half of) the volume of a unit sphere in 4 dimensions using a 3-dimensional integrand over the interval {[-1,1]}}, label = nsphere}
\begin{lstlisting}
$ mpiexec -n 5 ./hdintegrator.py \
>     --integrand integrands/N-sphere \
>     --dimensions 3 \
>     --min-extent -1 \
>     --max-extent 1
2.465467073965016 0.002690919078211793
\end{lstlisting}

We implement the integrand for functional integrals using the Monte Carlo integration algorithms of the GNU Scientific Library \cite{gsl}.
HDIntegrator communicates with the integrand via standard input and output in ASCII format.
Every line of input to the integrand consists of the number of samples and the extent of integration volume in every dimension.
Every line of output from the integrand consists of the result, absolute error and a suggestion in which dimension to split the subvolume in case convergence is not achieved.
Listing \ref{serial} shows an example invocation of the integrand for evaluating an integral and is essentially how the integrand is executed also by the parallel Python wrapper.

\lstset{language = bash, caption = {Example invocation and output of a integrand for calculating (half of) the volume of a unit sphere in 3 dimensions using a 2-dimensional integrand with $10^7$ samples over the interval {[-1, 1]}}, label = serial}
\begin{lstlisting}
$ echo 1e7 -1 1 -1 1 | ./integrands/N-sphere
2.093776991699273e+00 4.355051371078182e-04 1
\end{lstlisting}

\section{Single-time velocity correlation of Burgers equation}
\label{results}

We calculate the single-time velocity correlation of Burgers equation on a lattice of 10 spatial and 2 temporal points for which $L=2, M=10$ in (\ref{Gen2}) and transform the integration range from $\pm \infty$ to $\pm 1$.
We examine the convergence of the result by calculating the integral using different convergence criteria when the number of samples within each subvolume is doubled from $10^6$:
1) the absolute relative difference between results is at most between 2.5 \% and 8.5 \% (1.025 and 1.085).
2) the absolute difference between results is at most between 0.5 and 2.
3) the maximum absolute value of either result is at most between 0.5 and 2.

\begin{figure*}
\centering
\includegraphics[width = \textwidth]{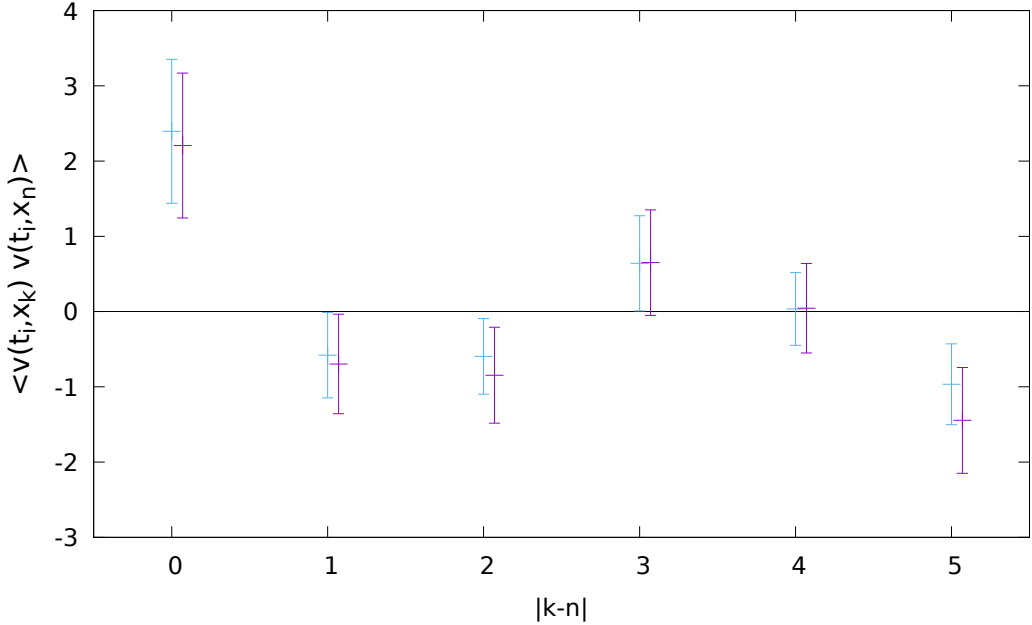}
\caption{
Normalized single-time velocity correlation of Burgers equation as a function of correlation distance.
Results with strictest convergence criteria are shown in cyan.
}
\label{fig:result}
\end{figure*}

Figure \ref{fig:result} shows the normalized single-time velocity correlation of Burgers equation in a 10x2 space and time lattice as a function of correlation distance in number of spatial points.
For each correlation distance, two results are shown for different convergence criteria used.
The results with strictest convergence criteria are shown in cyan.
Each value is normalized by calculating the integral with the velocity factors in (\ref{Gen2}) in front of the exponential set to 1.

\begin{figure*}
\centering
\includegraphics[width = \columnwidth]{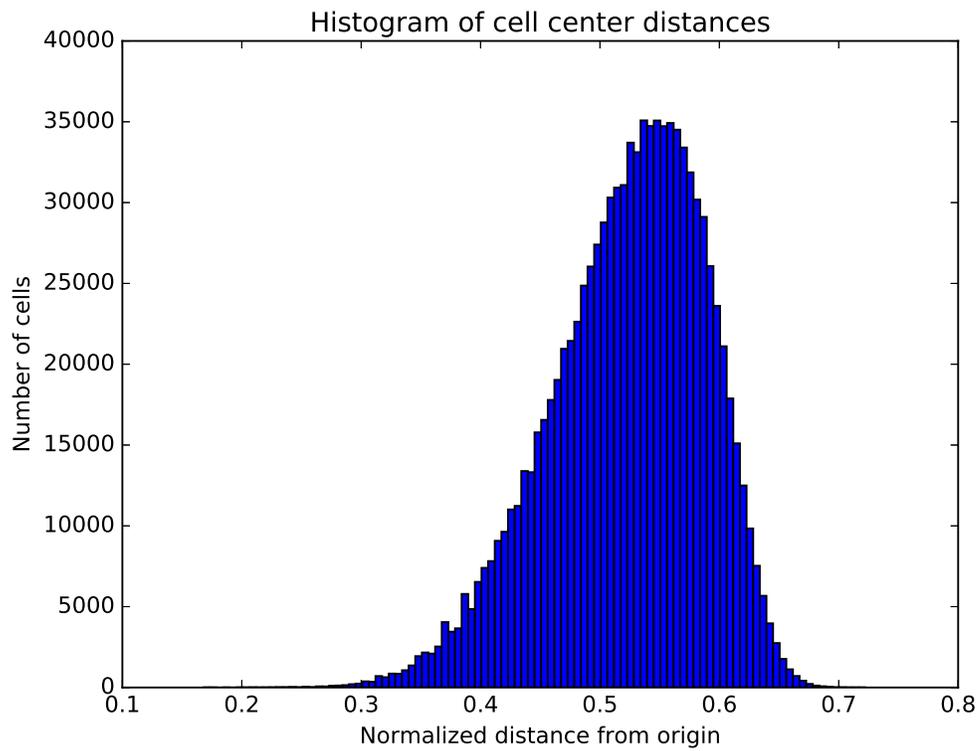}
\caption{
Histogram of number of centres of integration subvolumes as a function of normalized distance from origin.
Most subvolumes, and hence worst convergence, is concentrated on a spherical shell at a distance of approximately 0.55 from origin.
}
\label{fig:histogram}
\end{figure*}

Figure \ref{fig:histogram} shows a histogram of integration subvolume centres as a function of distance from origin for an integral of correlation distance of 5.
The distances from origin are normalized by $1/\sqrt{D}$ where D is the number of dimensions.
Integration proceeds to smaller and smaller subvolumes until convergence is reached and one can see that the largest number of subvolumes, and hence two worst convergence, is concentrated around a shell at a distance of approximately 0.55 from origin.

\section{Discussion}
\label{discussion}

The smallest computational resources required for obtaining one point in Figure \ref{fig:result} used approximately $500$ core hours of computational time and required a total of less than 10 GB of memory while the largest resources for one result required on the order of $10^4$ core hours and 100 GB of memory.
The required memory can most likely be decreased significantly by further optimizing its use in the parallel Python wrapper as well as switching from a Python implementation to e.g. C++.
Substantial gains in required computational time will probably require a different integrand from (\ref{Gen2}).
In this regard potential optimizations include switching from cartesian to spherical coordinates in order to better concentrate resolution where it is needed (cf.~Figure \ref{fig:histogram}) and/or directly calculating the Fourier spectrum of the single-time velocity correlation.
Utilizing GPUs and/or cloud computing is also an option worth exploring in the future.

\section{Conclusions}
\label{conclusions}

We present a method for modelling turbulence via numerical functional integration.
As a first step, we study single-time velocity correlation of Burgers' equation by transforming it into a functional integral that we solve with a parallel Python wrapper over the Monte Carlo integrators of GNU Scientific Library.

In contrast to direct numerical simulation, our method allows for simple parallelization of the problem as even a single integral can be evaluated independently over different subvolumes.
Here we evaluated each point in Figure \ref{fig:result} separately but in parallel using 140-280 Intel Haswell cores and at most our calculation used 1540 cores simultaneously of the Cray XC40 supercomputer installed at the Finnish Meteorological Institute.

The results show that, as a first step, our method is promising but there still remains substantial work in developing an optimized approach for integration that is most suitable to this particular problem.

\section*{Funding}
The work of IH was funded by the Emil Aaltonen foundation.

\bibliographystyle{tfnlm}
\bibliography{references}

\end{document}